\documentclass[twocolumn,preprintnumbers,amsmath,amssymb]{revtex4}
\usepackage{amsmath}    
\usepackage{amsmath,amssymb}
\usepackage{ascmac}
\usepackage{graphicx}   
\usepackage{verbatim}   
\usepackage{color}      
\usepackage{bm}
\usepackage{subfigure}  
\usepackage{hyperref}   
\usepackage[normalem]{ulem} 
\usepackage{siunitx} 

\begin{document}
\title{Thermal Molecular Focusing: Tunable Cross Effect of Phoresis and Advection}
\author{Tatsuya Fukuyama$^{1}$, Sho Nakama$^{1}$ and Yusuke T. Maeda$^{1,*}$}
\affiliation{$^1$Kyushu University, Department of Physics, Motooka 744, Fukuoka 819-0395, Japan}
\date{\today}

\begin{abstract}
The control of solute fluxes through either microscopic phoresis or hydrodynamic advection is a fundamental way to transport molecules, which are ubiquitously present in nature and technology. We study the transport of large solute such as DNA driven by a time-dependent thermal field in a polymer solution. Heat propagation of a single heat spot moving back and forth gives rise to the molecular focusing of DNA with frequency-tunable control. We developed a theoretical model, where heat conduction, viscoelastic expansion of walls, and the viscosity gradient of a smaller solute are coupled, and that can explain the underlying hydrodynamic focusing and its interplay with phoretic transports. This cross effect may allow one to design a unique miniaturized pump in microfluidics. 
\end{abstract}

\maketitle
In 1951, the seminal work by G. Taylor has shown that the pumping in a viscous fluid can be driven by an undulating infinite sheet at a low Reynolds number, proposing that the flexible object can put into motion of a viscous fluid with a finite surface disturbance\cite{taylor}. To date, a variety of analytical and numerical models for pumping in a viscous fluid have been developed \cite{jaffrin}, and the pumped flow of complex fluids has gained considerable attention from a fundamental perspective. A peristaltic flow through periodic contact compression using electromechanical\cite{schmidt} or opto-mechanical stresses\cite{cuennet} has been shown to be microfluidic modulators. In particular, a hydrodynamic force offers a versatile method for rapid mixing\cite{austin}, particle trapping, and assembly\cite{anish} in a confined geometry with dimensions of tens or hundreds of micrometers, which has revolutionized fluid mechanics and soft-matter physics at the small scale as the core of nano- to microfluidic devices\cite{whitesides}.  

To open up a new avenue for the use of hydrodynamic forces, explorations beyond conventional elements are challenging. On the one hand, light-driven advection of particles has been demonstrated by using an infrared laser focusing. When a hot spot in a focused laser moves at a constant speed in a highly viscous solution, a net fluid flow occurs, owing to the coupling of thermally reduced viscosity and fluid compressibility\cite{yariv}\cite{weinert}\cite{pal}. Although this effect can convey the particles, to be extended for the trapping of molecules, one requires a well-designed stagnation point under complex streamlines. On the other hand, microscopic phoretic transports such as thermophoresis\cite{jonas}\cite{wurger}\cite{marco}\cite{lin} or diffusiophoresis\cite{diffusio}\cite{abecassis}\cite{palacci}\cite{selva}\cite{florea}\cite{shin}\cite{shi}, which is the transport along a gradient of temperature or concentration of a smaller solute respectively, is expected to be a versatile mean of molecular manipulation. Thermophoresis depletes a high concentration of a solute from a hot region and builds its concentration gradient. In such a solution, another solutes of larger size experience both thermophoresis and diffusiophoresis as a secondary effect. The balance of two phoretic motions allows one to control the direction and magnitude of the transport velocity to trap molecules\cite{sano}\cite{mae1}\cite{mae2}. Indeed, the phoretic manipulation has an advantage, that is, less material dependence than conventional optical tweezers, because the driving force behind it is a surface slip velocity, which is essentially never affected by the material properties\cite{mae3}\cite{mae4}. However, the balance between counter-acting transports has to be suitably adjusted by changing the temperature difference or solute concentration under the initial conditions, which could be a fundamental limitation. Conventional methods thus have exhibited particular advantages and limitations. Hence, further advances in our understanding regarding the interplay among phoretic transports and hydrodynamics are needed. 

In this Letter, we report thermal molecular focusing, where microscopic phoretic transports are coupled to advective flow tunable by a time-dependent thermal field in a polymer solution. A moving heat source is able to pump the advective flow based on the effects arising from the heat conduction, thermally reduced viscosity of polymer solution, and viscoelastic compression from the boundary. The cross effect among two types of phoretic transports and thermally induced flow allows one to control molecular manipulation without a suitable adjustment of the external fields. Strikingly, time-delay involved in viscoelastic compression is important in accounting for both molecular focusing and frequency-tunable control. Our finding thus impacts to a broad-class of soft viscoelastic materials, from nematogen microfluidics to epithelial cell sheet, that is out of equilibrium.

\textit{Experimental setup.---}  We built the following experimental setup\cite{supporting1}. A thin chamber with a thickness of \SI{25}{\micro\meter} and width of \SI{800}{\micro\meter} was made using polydimethyl siloxane (PDMS), and within which a bulk solution was confined. Such a small thickness suppresses the thermal convection during laser irradiation. An infrared laser (Furukawa electronics, \SI{1480}{\nm} wavelength), focused using a 20$\times$ objective lens, was used to build a temperature field of $\Delta T(x,y)=T(x,y)- T_0$, where $T_0=24\pm$\SI{0.1}{\degree C} is the temperature at infinity. A typical temperature gradient is $\nabla T$=\SI{0.08}{K/\micro\meter}, and the maximal difference $\Delta T$=9.6 K\cite{mae1}\cite{mae2}\cite{mae4}. The laser spot was steered by using galvo-scan mirrors (Cambridge Technologies)\cite{mae3}. We used DNA of 4.3 kbp as a large solute (gyration radius $a$$\approx$\SI{0.1}{\micro\meter}) with a concentration of 0.01\textit{wt}\%. The DNA was then dissolved in a solution of 5.0\textit{wt}\% polyethylene glycol 20000 (PEG) as a smaller solute (gyration radius $R_g^p \approx$ \SI{2.5}{\nano\meter}). The heat source was moved back and forth along a line $-L \leq x \leq L$ with $2L$=\SI{209}{\micro\meter} at a speed of $u_l$. The DNA was then visualized using the SYBR Gold dye and its concentration was measured using an epi-fluorescent microscope (Olympus, IX73).

\textit{Thermal molecular focusing.---}  We first present the basic phoretic transports in gradients of  temperature and solute concentration, at a fast moving hot spot ($u_l\sim$\SI{10}{\milli\meter\per\second}) along the linear path. If we take a large volume fraction of solute A in a temperature gradient, a gradient of solute concentration can be built. In such a solution, another solute B of a very small volume fraction will be displaced out of the hot region whereas the diffusiophoretic transport caused by the gradient of solute A tends to bring it back. Increasing the concentration of A from zero, the local accumulation of B can be observed, and is finally trapped at the hot spot\cite{sano}\cite{mae1}\cite{mae3}. Consistent with this type of mechanism, a nearly stable gradient of PEG is built owing to the occurrence of thermophoresis and the PEG concentration gradient as a secondary effect of thermophoresis accumulates the DNA uniformly within the heated region along the scanning path (FIG. 1(a))\cite{mae1}\cite{mae3}. In addition, at a slower velocity ($u_l\sim$\SI{1}{\micro\meter\per\second}), the DNA is uniformly trapped along the path of laser scanning as long as the diffusion of the DNA is much slower than the period of heat stimulus. Although these molecular trappings at two particular cases can be explained from conventional phoretic transports, a new behavior was observed at the intermediate velocity ($u_l\sim$\SI{e2}{\micro\meter\per\second}). The DNA was accumulated not along the path of the moving hot spot, but at its mid-point, and was finally focused (FIG.1 (a) and (b)). The amount of focused DNA was controlled in a velocity-dependent manner, implying time-dependent modulation for the direction of transport. This finding motivated us to explore two questions: First, what effect can give rise to molecular transport based on the speed of a moving hot spot? Second, what is the underlying mechanism behind the focusing of DNA through a dynamic thermal gradient? 
\begin{figure}[tb]
 \begin{center}
  \includegraphics[width=90mm]{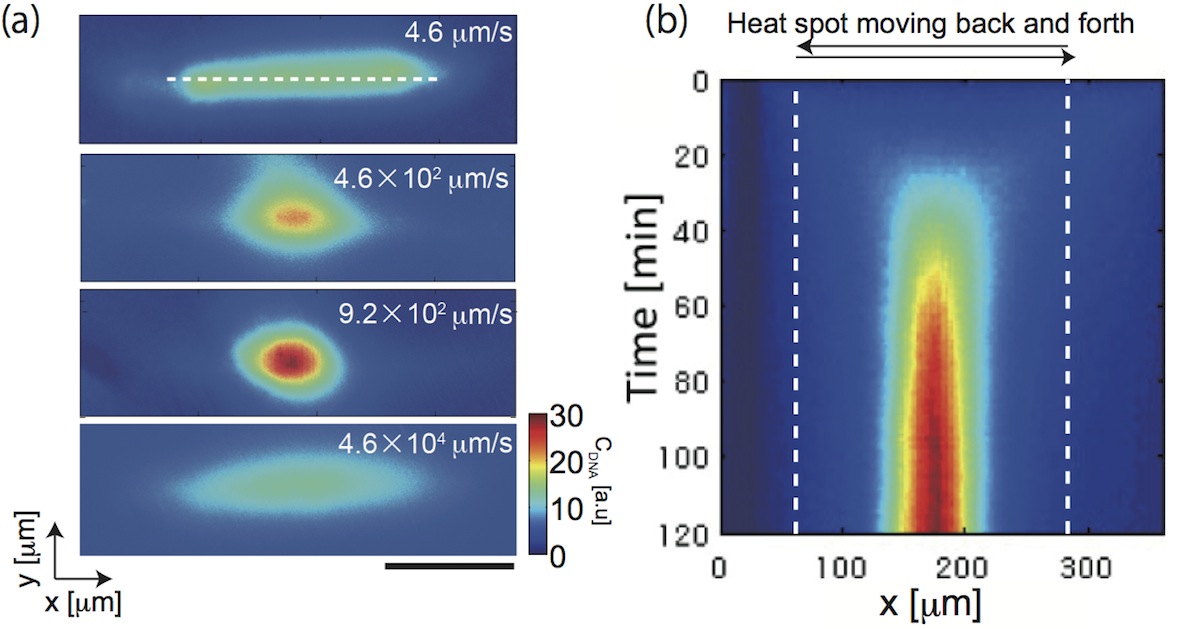}
 \end{center}
 \caption{Experimental demonstration of tunable molecular focusing of DNA using a moving hot spot in a polymer solution. (a) DNA trapped by the temperature gradient moving back and forth at various speed $u_l$=4.6, \num{4.6d2}, \num{9.2d2}, and \SI{4.6e4}{\micro\meter\per\second} (from top to bottom). The white dashed line is the path of a moving laser spot. Scale bar: \SI{100}{\micro\meter}. (b) Kymograph of molecular focusing of DNA at $u_l$=\SI{9.2e2}{\micro\meter\per\second}. The white dashed lines are the edges of the path of a moving laser spot}\label{fig1}
\end{figure}

\textit{Microflow.---}  It has been shown that the onset of a net fluid flow is driven by a moving heat source\cite{weinert}\cite{pal}. When a localized hot spot moves in a highly viscous solution (e.g., 80\textit{wt}\% glycerol) confined within thin solid substrates (a height of a few microns), thermal expansion of the viscous fluid induces extensile and contractile flows at the front and rear edges of the hot spot, respectively. Although the isotropic viscosity prohibits the onset of net flow, thermally reduced viscosity in fluids enlarges the extensile flow at the front edge and the contractile flow at the rear edge. This imbalance eventually creates a net flow opposite to the motion of the hot spot. This thermo-hydrodynamic flow at microscale (microflow) could be relevant to the observed thermal molecular focusing. However, this type of advection becomes zero after canceling each other out when the hot spot moves back and forth. To address the underlying mechanism in thermal molecular focusing, our key idea is to combine distinct disciplines, those are hydrodynamics, viscoelastic mechanics, and heat conduction, as formulated below. 

Herein, we propose a model that leads to both molecular focusing and its tunable control in a moving heat source\cite{supporting1}. A semi-dilute polymer solution is enclosed in a thin chamber with a deformable PDMS wall. Its viscosity, elastic modulus, and characteristic relaxation time are $\eta^w$, $E$, and $\tau=\eta^w/E$, respectively. A moving heat source propagates in one direction with periodic boundary conditions (FIG. 2(a)). The walls in top and bottom deform upon the transferred heat at time $t=0$, and thereafter the enclosed fluid is compressed. The strain relaxation of the deformed wall after $\Delta t$ is set using $\epsilon_{\tau}(\Delta T, \Delta t)=\Delta V_{\tau}/V_0\approx\gamma (1 - e^{- \Delta t/\tau}) \Delta T$ where $\gamma=\frac{1}{V_0}\frac{dV}{dT}$ is the thermal expansion coefficient of the wall substrate. In addition, thermally reduced viscosity becomes significant owing to thermophoresis of polymer solute, which is expressed by $\beta$=$\frac{1}{\eta}\frac{d \eta}{dT}$=$\frac{1}{\eta}(\frac{\partial \eta}{\partial T}+\frac{\partial \eta}{\partial c}\frac{\partial c}{\partial T})$, including thermophoresis (second term) (FIG. 2(b)). By averaging over a single period of thermal stimulation $1/f_l$, the microflow is
\begin{equation}
\bm{u_{ve}} = - \bm{u_l}\frac{\beta \Gamma_{\tau}}{2}  (\Delta T)^2 \label{microflow1}
\end{equation}
where $\Gamma_{\tau}=\gamma (1-e^{-1/f_l\tau})$ is thermal \textit{viscoelastic} coefficient of the wall\cite{supporting1}. 

For a PEG solution, thermophoretic depletion allows one to have a more explicit form of $\beta$. In a temperature gradient of $\nabla T$, the density flux $J^p$ of a PEG solute is given by $J^p$=$-D^p (\nabla c^p + c^p S_T^p \nabla T)$ where $c^p$ is the PEG concentration, $D^p$ is the diffusion coefficient of PEG, and $S_T^p$ is the Soret coefficient of PEG defined as $D_T^p/D^p$ with the thermal diffusion coefficient of PEG of $D_T^p$\cite{chan}($D^p$=\SI{58}{\micro\square\meter\per\second} and $S_T^p$=\SI{8.89e-2}{\per\kelvin} in this study). It is known that the viscosity logarithmically increases with $c^p$\cite{holyst}, and one yields $\beta=\beta_0+\beta_1 c_0^{p} S_T^{p}$\cite{supporting2}. Substituting $\beta$, Eq.(\ref{microflow1}) is rewritten as
\begin{equation}
\bm{u_{ve}} = - \bm{u_{l}} \frac{(\beta_0 + \beta_1 c_0^p S_T^p)\Gamma_{\tau}}{2}(\Delta T)^2. \label{microflow2}
\end{equation}

\begin{figure}[t]
 \begin{center}
  \includegraphics[width=85mm]{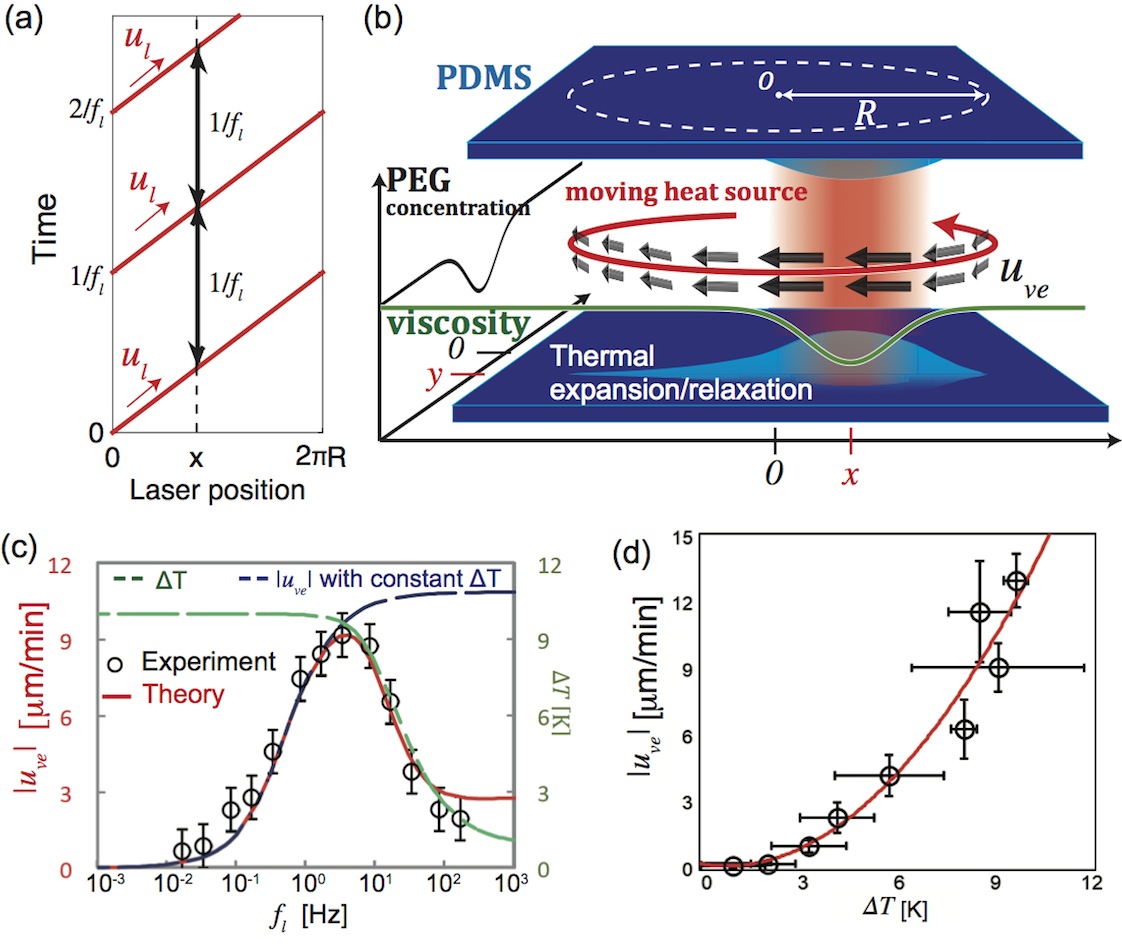}
 \end{center}
 \caption{Frequency-dependent flow driven by a moving hot spot. (a) Time-dependent thermal field created using a propagating heat source. A single heat source moves in a unidirectional manner under periodic boundary conditions. (b) Schematic image of heat-induced advection, through the coupling of thermal viscosity reduction and thermal viscoelastic expansion, by a moving heat source. (c) Frequency dependence of $|\bm{u}_{ve}|$ as a function of $f_l$ from \SIrange{1.65e-2}{1.65e2}{\hertz}. The blue dashed curve represents the frequency dependence without considering the reduced temperature difference at a higher frequency (assuming a constant $\Delta T$=\SI{9.6}{\kelvin}). The green dashed curve shows the maximum temperature difference $\Delta T$ calculated from Eq.(3) using the following parameters\cite{erickson}\cite{wedershoven}: $C_v$=\SI{4.2e-12}{\joule\per\micro\cubic\meter\per\kelvin}), $\lambda_h$=\SI{6.1e-7}{\watt\per\micro\meter\per\kelvin}), and $h$=\SI{5.0e-10}{\watt\per\micro\cubic\meter\per\kelvin}). (d) Quadratic increase of the microflow as a function of temperature difference. The frequency of the thermal stimulus was set at $f_l$=\SI{3.2}{\hertz}.}\label{fig2}
\end{figure}
To experimentally test Eq.(\ref{microflow2}), we created a moving hot spot that rotates in one direction along a circular path\cite{supporting1}. The radius of the circular path is $R$=\SI{150}{\micro\meter} and the frequency of thermal stimulation is $f_l$=$u_l/2\pi R$. We measured the spatial profile of local microflow through particle image velocimetry  \SI{3.0}{\micro\meter} silica beads as tracer particles. FIG. 2(c) shows $u_{ve}$ for a broad range of $f_l$=\SIrange{0.2}{200}{\hertz}. For \SIrange{0.2}{3.2}{\hertz}, the microflow gradually increases to a peak of $|u_{ve}|\approx$\SI{10}{\micro\meter}/min, showing good agreement with theoretical calculation at $\tau=1.2$ sec. Furthermore, we found that $u_{ve}$ increases quadratically with $\Delta T$ as experimentally confirmed for $\Delta T$=\SIrange{1.0}{9.6}{\kelvin} at $f_l$=\SI{3.2}{\hertz} (FIG. 2(d)). This quadratic increase evidently proves the multiplicative coupling of thermo-viscoelastic expansion and reduced viscosity after thermophoretic depletion of PEG.

\begin{figure}[t]
 \begin{center}
  \includegraphics[width=85mm]{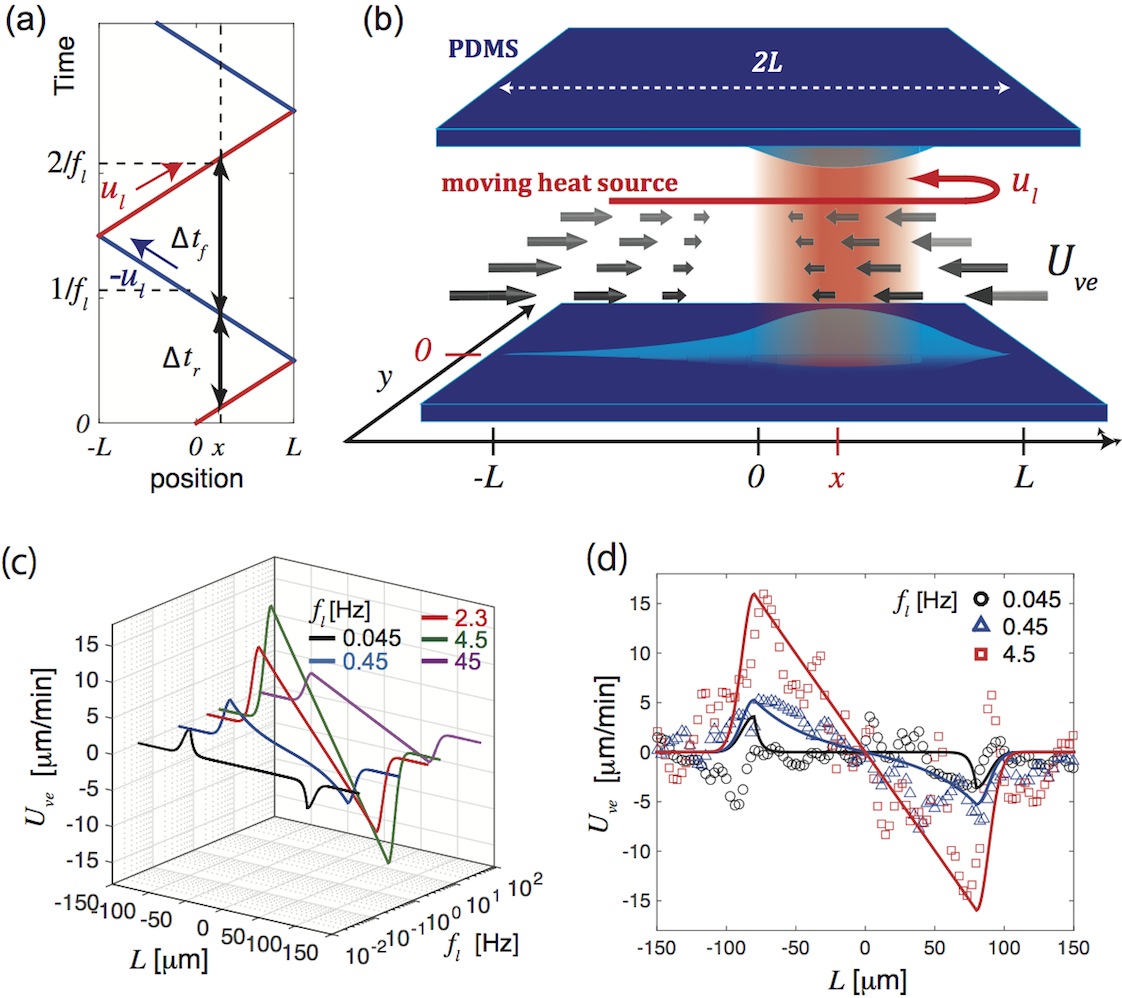}
 \end{center}
 \caption{Hydrodynamic focusing in a hot spot moving back and forth. (a) A single heat source moves in back and forth in direction. It turns the direction of motion in the opposite direction at the edges $x=\pm L$. (b) Schematic image of LHF through the coupling of thermal viscosity reduction and thermal viscoelastic expansion in a hot spot that moves in back and forth. (c) An analytical plot of the flow velocities, numerically calculated from Eqs.(\ref{heat}) and (\ref{LHF}) with the same parameters. (d) The profiles of LHF at $f_l$=\SI{4.5e-2}{\hertz}(black), \SI{4.5e-1}{\hertz}(blue), 4.5 Hz (red). The solid lines are numerical calculations using same parameter set.}\label{fig3}
\end{figure}

An intriguing observation is the frequency dependence of $u_{ve}$. FIG. 2(c) shows that $u_{ve}$ has a peak at an intermediate frequency of $f_l$=3.2 Hz, and then decreases to \SI{1.0}{\micro\meter\per\minute} at $f_l \geq 1/\tau$. A hot spot moving at higher frequency was also unable to enhance DNA accumulation (FIG. 1(a)), implying that the velocity-dependence of molecular focusing reflects the resonant-like profile of $u_{ve}$. Eq. (\ref{microflow2}) indicates that $u_{ve}$ becomes insensitive to $f_l$ at a frequency beyond the lower threshold $f_l^{*}\approx1/\tau$, according to its frequency dependence of $f_l(1-\exp[-1/(f_l\tau)])$ when assuming constant $\Delta T$. The observed decay of the microflow points out the contribution from the dynamics of the heat flux. 

PDMS is a thermal insulation material with low coefficient of heat transfer and enables varying $\Delta T$ in a frequency dependent manner. Heat conduction across the wall is evaluated based on the thermal diffusion equation in a two-dimensional space $(x,y)$, where a hot spot moves at a constant speed $u_{l}$ along $x$ axis:
\begin{equation}
C_{v} \frac{\partial (\Delta T)}{\partial t} - \lambda_h \nabla^2 (\Delta T)= P - h \Delta T, \label{heat}
\end{equation} 
where $C_{v}$ and $\lambda_h$ are the heat capacity and the heat diffusion coefficient of water respectively, $P$=$P_0 \exp[-((x-u_l t)^2+y^2)/(2b^2)]$ with a spot radius of $b$=\SI{7.5}{\micro\meter} is thermal energy from the laser spot, and $h$ is the thermal transfer coefficient that represents heat sink toward PDMS from the solution\cite{erickson}\cite{wedershoven}. $\Delta T$ also has a frequency dependence as $\Delta T$ starts to be reduced at a higher frequency $f_l \geq 2 \pi h/C_v \approx$ \SI{19}{\hertz}. Theoretical curve calculated from Eqs.(\ref{microflow2}) and (\ref{heat}) agrees well with experiment (FIG. 2(c)).

\textit{Local hydrodynamic focusing.---}  We next study the link between microflow and thermal molecular focusing (FIG. 3(a)). A heat source moves back and forth along a line ($-L \leq x \leq L$, $y=0$, $2L$=\SI{160}{\micro\meter}) at a velocity of $\bm{u_l}$=$u_l \bm{e_x}$, where $\bm{e_x}$ is the unit vector on the $x$ axis. The polymer solution is exposed to thermal stimuli with two different time intervals of either $\Delta t_f$=$2(L-x)/u_l$ in the forward direction or $\Delta t_r = 2(L+x)/u_l$ in the backward direction; this corresponds to $\Delta V_{\tau,f}/V_0$=$\gamma(1-e^{-\Delta t_f/\tau})$ and $\Delta V_{\tau,r}/V_0$=$\gamma(1-e^{-\Delta t_r/\tau})$ for a thermal viscoelastic coefficient respectively. By summing forward and backward microflows at $x$ as the average over time, we have 
\begin{equation}
\bm{{U}_{ve}}(x) = - \bm{U} \sinh \Bigl[\frac{x}{f_l \tau L} \Bigr] (\beta_0 + \beta_1 c_0^p S_T^p) \gamma(\Delta T)^2,\label{LHF}
\end{equation}
where $f_{l}$=$u_{l}/2L$, defined as $1/f_{l}$=$(\Delta t_f + \Delta t_r)/2$, is the effective frequency of a thermal pulse, and $\bm{U}$=$2L f_{l} \exp[-1/(f_{l}\tau)] \bm{e_x}$ is the frequency-dependent fluid velocity. Remarkably, the hyperbolic sine function in Eq.(\ref{LHF}) represents the spatial profile of the microflow, which immediately leads to $U_{ve}(x)\propto -x$ for a small $x$. The microflows from both ends face one another and then collide at the midpoint (FIG. 3(b)), that is local hydrodynamic focusing (LHF). As shown in FIG. 3(c), $\bm{U}_{ve}(x)$ at various velocities $u_l$ clearly exhibits a flow oriented toward the midpoint from both edges. We experimentally verified the LHF of Eq. (\ref{LHF}) in a 5.0\% PEG solution by forcing the laser spot back and forth along the line with $2L$=\SI{160}{\micro\meter}. The proportional change of flow velocity to the distance from the midpoint is consistent with theoretical result (FIG. 3(d)). 

\textit{Interplay of diffusiophoresis and LHF.---}  Key finding of LHF motivated us to investigate whether the interplay among phoretic transports (thermophoresis and diffusiophoresis) and LHF underlies tunable molecular focusing of DNA. We define the density flux of DNA by $\bm{J}=\bm{J}_{diff}+\bm{J}_{Tp}+\bm{J}_{Dp}+\bm{J}_{ve}$, where normal diffusion (first term), thermophoresis (second), diffusiophoresis (third), and LHF (fourth) are considered, and the net flux $\bm{J}$ is 
\begin{equation}
\bm{J} = -D (\bm{\nabla} c + c S_T \bm{\nabla} T) + c\bm{U}_{Dp} + c\bm{U}_{ve}, \label{flux}
\end{equation}
where $c$ is the concentrations of DNA, $D$ and $S_T$ the diffusion coefficient and the Soret coefficient of DNA respectively ($D$=\SI{2.89}{\micro\meter}$^2$/s and $S_T$=0.38 K$^{-1}$). Because diffusiophoresis of DNA in 5.0\textit{wt}\% PEG solution becomes dominant rather than thermophoresis\cite{sano}\cite{mae1}, the balance between LHF of $\bm{U}_{ve}$ and diffusiophoresis of $\bm{U}_{Dp}$ decides where DNA is accumulated. $\bm{U}_{ve}=(U_{ve}, V_{ve})$ conveys DNA to a stagnation point at the center $(x,y)=(0,0)$ (FIG. 4(a)) whereas diffusiophoresis $\bm{U}_{Dp}=(U_{Dp}, V_{Dp})$ captures DNA along $x$ axis (FIG. 4(b)). Accordingly, their interplay focuses DNA at the mid-point. 

\begin{figure}[tb]
 \begin{center}
  \includegraphics[width=88mm]{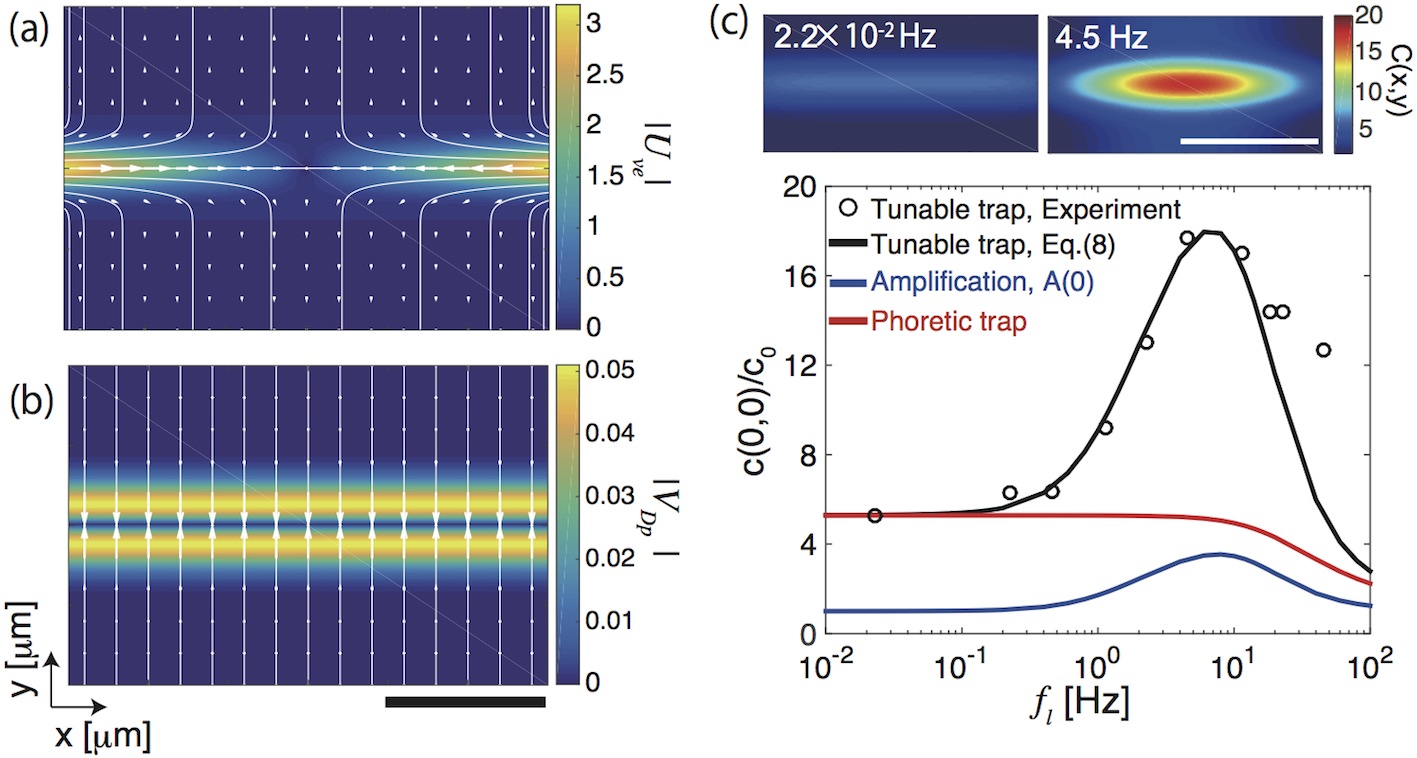}
 \end{center}
 \caption{Tunable, thermal molecular focusing of DNA by the interplay between LHF and phoretic transports. (a) A color-map of flow speed in the $x$ direction, $|U_{ve}|$, is simultaneously shown with its streamline (white) in a two-dimensional plane. (b) A color-map of the diffusiophoretic flow in the $y$-direction, $|V_{Dp}|$, is also shown with its streamline (white). Scale bar: \SI{100}{\micro\meter}. (c) Frequency tunable focusing of DNA. The black circle represents normalized amount of DNA in experiment, $c(0,0)/c_0$, and the black line is the numerical calculation of Eq. (\ref{DNA}). The blue line represents $A(0)$ as a function of $f_l$, and red shows the accumulated large solute solely from thermophoresis and diffusiophoresis, $\exp[-S_T \Delta T + V'\Delta c^p]$. Inset shows a color-map of $c(x,y)$ obtained by numerical calculation at \SI{2.2e-2}{\hertz} (left) and \SI{4.5}{\hertz} (right).}\label{fig4}
\end{figure}

We next solve Eq. (\ref{flux}) at the steady state ($\bm{J}$=0) to theoretically analyze molecular focusing of DNA in more detail. If the recovery of depleted PEG along the $x$ axis, which is $2bL/D^p\approx$20 sec, takes longer than the period of a moving hot spot $1/f_l$, the gradient of PEG can be assumed to stable. As for $f_l\geq0.1$ Hz, we consider $\partial_x c^p$=$0$ and $\partial_y c^p$ no longer depends on time, i.e. $c^p(x,y)\approx c^p(y)=c_0^p \exp[-S_T^p \Delta T(y)]$.  Diffusiophoresis transports DNA perpendicular to the laser path ($y$=0) with $\bm{U}_{Dp}=(0, V_{Dp})$ and $V_{Dp}=\frac{k_B T}{3 \eta}\lambda^2 (S_T^p - \frac{1}{T})c^p(y)\partial_y T$\cite{sano}\cite{mae1}. Moreover, because $V_{Dp}$ in 5.0\% PEG with $\nabla T \approx 0.1$ K/$\mu$m is much larger than $V_{ve}$, we can simplify the density flux of DNA as
\begin{eqnarray}
J_x &=&-D (\nabla_x c + cS_T \nabla_xT) + cU_{ve},\label{eq:x-transport}\\
J_y &=&-D (\nabla_y c + cS_T \nabla_yT) + cV_{Dp}.\label{eq:y-transport}
\end{eqnarray}
 By solving Eqs. (\ref{eq:x-transport}) and (\ref{eq:y-transport}) at the steady state $J_x=J_y=0$, one obtain the concentration of focused DNA as 
\begin{equation}
c(x,y)=c_0 \exp\Bigl[-S_T \Delta T + V' \Delta c^{p} +\frac{1}{D} \int_{-\infty}^{x} U_{ve}(x') dx'\Bigr], \label{DNA}
\end{equation}
where $\Delta c^{p}=c_0^p-c^p$ is the depleted amount of PEG, and $V'=2\pi a \lambda^2$ the effective volume involved in diffusiophoresis with a depletion layer of thickness of $\lambda$ ($\lambda$ is comparable to $R_g^p\approx$2.5 nm). FIG. 4(c) shows the normalized concentration of focused DNA of $c(0,0)/c_0$. This analytical result agrees well with the experiment entirely for $f_l$=0.02 to 200 Hz. 

The amplification rate of molecular focusing, which is defined as $A(x)=\exp[1/D \int_{-\infty}^{x} U_{ve}(x') dx']$, i.e. $\ln A(x)\cong\beta \gamma \frac{f _l \tau L U_{l}}{D} \bigl(\cosh [1/(f_l\tau)]-\cosh[(x/L)/(f_l\tau)] \bigr)(\Delta T)^2$, tells the strategy to extend this mechanism toward various-sized objects. Here, $\ln A(0) \approx \beta \gamma L^2 f_l /D (\Delta T)^2$ is yielded as the approximated maximum of the amplification rate close to a inflection point $f_l^* \tau =1$. If the size of a trapped molecule is small (large $D$), the length of $L$ has to be enlarged so as to keep $L^2/(\tau D)\gg1$. Thermal molecular focusing is thus applicable for various sized materials.

\textit{Conclusion.---} In this Letter, we presented the frequency-tunable focusing of molecules that results from the interplay between LHF, diffusiophoretic and thermophoretic transports in a polymer solution. This simple, versatile method controls molecular transport owing to time-dependent viscoelastic compression, and shows quantitative agreement with theoretical model. A suitable adjustment of the external fields, which is a prerequisite in the existing method, is no longer required. The frequency tunable trapping thus offers new perspectives for the microscale manipulation of particles, which may also enable the design of miniaturized pump in opto-microfluidics\cite{schmidt}. For this aim, gel-based substrate with a large $\tau$ can speed up a faster net flow. Moreover, the viscoelastic bulk fluids, e.g., nematic liquid crystals, may extend the revealed mechanism to three-dimensional architectures that will further enhance the development of peristaltic nematogen microfluidics\cite{cuennet}\cite{kim}\cite{skarabot}. The conceptual advance from present study is not limited to only practical applications but may be relevant to active fluids. Autonomous viscoelastic compression coupled to the mobility could be involved in collective cell migration\cite{yabunaka}\cite{aoki}, so that extensive work may bridge the gap between physics and biology in future studies.

\textit{Acknowledgements}
We thank Y. Kimura, Z. Izri, R. Sakamoto, N. Yoshinaga for critical reading. This work was supported by PRESTO grant (No.11103355, JPMJPR11A4) from Japan Science and Technology Agency, JSPS KAKENHI (16H00805: "Synergy of Fluctuation and Structure" and 17H00245 Grant-in-Aid for Scientific Research (B)) from MEXT, and HFSP research grant (RGP0037/2015).

\newpage

\section*{Supplemental information for \\Thermal Molecular Focusing: Tunable Cross Effect of Phoresis and Advection}
\author{Tatsuya Fukuyama$^{1}$, Sho Nakama$^{1}$ and Yusuke T. Maeda$^{1,*}$}
\affiliation{$^1$Kyushu University, Department of Physics, Motooka 744, Fukuoka 819-0395, Japan}
\affiliation{$^*$corresponding address: ymaeda@phys.kyushu-u.ac.jp}
  \maketitle

\newcommand{\beginsupplement}{%
        \setcounter{table}{0}
        \renewcommand{\thetable}{S\arabic{table}}%
        \setcounter{figure}{0}
        \renewcommand{\thefigure}{S\arabic{figure}}%
     }
     
     \beginsupplement

\section{Experimental details}
\subsection{Optical setup}
The solution was entrapped in a chamber of \SI{25}{\micro\meter} thickness and \SI{800}{\micro\meter} in diameter made by standard soft lithography techniques described below. This small thickness suppresses the onset of thermal convection during laser irradiation. The chamber was viewed with an epifluorescence microscope (Olympus, IX73) with the stable excitation light source (Lumen Dynamics, XLED1). The temperature of the microscope stage was kept at room temperature (=24.0 $\pm$\SI{1.0}{\celsius}). 
\begin{figure}[b]
 \begin{center}
 \includegraphics[width=80mm]{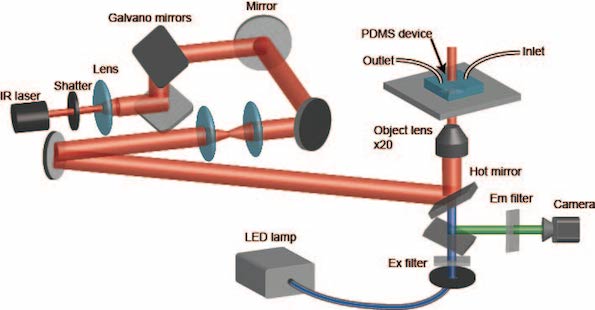}
 \end{center}
 \caption{Schematic illustration of optical setup used in this study}\label{figS1}
\end{figure}
Temperature gradient was built by focusing infrared laser (FIG.S1). Photons of \SI{1480}{\nano\meter}wavelength are efficiently absorbed in water. The temperature difference {\it in situ}, $\Delta T(x,y) = T(x,y) - T_0$, was measured by calibrating the reduced fluorescent intensity of temperature-dependent fluorescein (2'-7'-bis(carboxyethyl)-5(6)-carboxyfluorescein, BCECF, Molecular probes). For this calibration we measured the fluorescent intensity of fluorescein at various temperatures from $T$=20 to \SI{50}{\celsius} by fluorescent spectrometer with a temperature control unit and then draw the curve for temperature calibration. We eventually found that the fluorescent intensity was decreased at the rate of -1.8\%/K. 
Pump photodiode, the temperature controller and current controller were purchased from Furukawa DENKO (FOL1435R50-317, Furukawa Electronics) and ILX Lightwave, respectively. The focused laser was deflected by a set of two galvo mirrors (Cambridge technologies). Other optical setup were purchased from Thorlabs. Infrared laser focusing was carried out through a 20$\times$ objective lens with long working-distance (NIKON).

\subsection{Microfabrication}
The microfluidic device made of silicone elastomer (PDMS, Sylgard 184, Dow Corning) with a PDMS-coated glass slide were filled with the water solution (FIG.S2). Patterned surface of this device was made from the mold of SU-8 3025 photoresist. SU-8 photoresist was patterned with conventional UV photolithography method. The patterned surface of SU-8 was transferred to PDMS chip by casting uncured PDMS mixed with curing agent and then cured for 1 hour at \SI{75}{\celsius}. After the cutting PDMS chip with a scalpell, the PDMS chip was strongly bonded on the PDMS-coated glass slide by having their surfaces hydrophilic by plasma-gun surface treatment for 30 sec and then cured PDMS again by heating at \SI{90}{\celsius} for 1 hour . The inlet and outlet of PDMS device were connected with thin PEEK tubes from the pressure-regulated microfluidic pump (MFCS flow system, Fluigent). 

\begin{figure}
 \begin{center}
 \includegraphics[width=80mm]{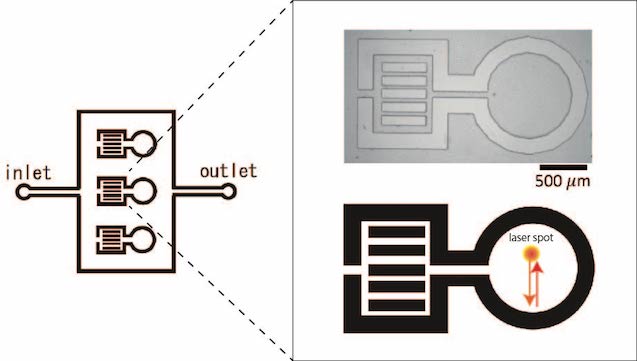}
\end{center}
 \caption{The design of the microfluidic device used in this study}\label{figS2}
\end{figure}

\subsection{Chemical reagents}
To demonstrate the trapping and focusing of molecules, we used plasmid DNA (pBR322 with 4.3 kbp. Its gyration radius is \SI{0.1}{\micro\meter}). The plasmid DNA was purified from {\it E.coli} bacteria by using conventional method. Purified DNA was stained by SYBR Gold dye (S11494, Molecular Probe) in order to quantitatively measure local concentration of DNA. The purified DNA was dissolved in Tris-HCl (pH 7.2) and 50 mM EDTA solution. The polymer dissolved in an aqueous solution is polyethylene glycol 20000 (Alfa Aesar) at 5.0\% weight per volume in Tris-HCl buffer solution. In order to avoid both evaporation of solvent content and bubble formation, the PEG solution was kept flowing continuously outside of the area for observation in the PDMS device (FIG. S2). 

\subsection{Image analysis}
As explained in the main text, particle image velocimetry (PIV) is performed for the visualization of flow field. The tracer particle for PIV was the silica beads with \SI{3.0}{\micro\meter} in diameter (Micromod, sicastar) and its mass fraction was set at 0.1\textit{wt}\% in 5.0\textit{wt}\% PEG solution. Time-lapse movie was taken by the interval of 3 sec and the obtained velocity field was analyzed by using Image J. The azimuthal velocity along the circular path of scanned laser was plotted as the frequency of thermal stimulus.
The frequency-dependence of fluid flow is observed by using the tracer particle of \SI{2.0}{\micro\meter} of charged polystyrene beads (Molecular probes), and both tracer particles show the speed of fluid flow  $|u_{ve}|\sim$\SI{10.0}{\micro\meter}/min at the peak $f_l=3.2$ Hz. 

To build the moving temperature gradient in one direction at a constant speed $u_{l}$, the path of a moving laser spot draws the circle with a radius $R$=\SI{150}{\micro\meter} and its tangential vector $\bm{e}_{\theta}=(-\sin \theta, \cos \theta)$ in polar coordinates whose origin is the center of circular path of laser scanning. The laser spot rotates in counter-clockwise direction at constant angular velocity of $\omega= 2 \pi f_l$, which also yields the frequency of a hot spot defined as $f_l=u_l/(2 \pi R)$. We measured fluid streaming by particle image velocimetry (PIV) in the solution of 5.0\% PEG that contains \SI{3.0}{\micro\meter} silica beads as tracer particles and then analyzed velocity field $\bm{u_{ve}}(r,\theta)$ by using its product to tangential vector averaged over angle, $\bar{u}_{ve}(r)=\langle \bm{u_{ve}}\cdot \bm{e_t} \rangle_{\theta}$ in order to extract primary component driven by thermal viscoelasticity. We found that the pumping of fluid flow was occurred in the opposite direction to the temperature wave. FIG.S3 shows the plot of $\bar{u}_{ve}(r)$ in the axis of radial distance $r$. It can be found that the fluid motion has a peak at $r=R$, which is the path of the moving laser spot, for all frequencies and the decay of $\bar{u}_{ve}(r)$ occurs over radial length symmetrically. 
 
 \begin{figure}
 \begin{center}
  \includegraphics[width=80mm]{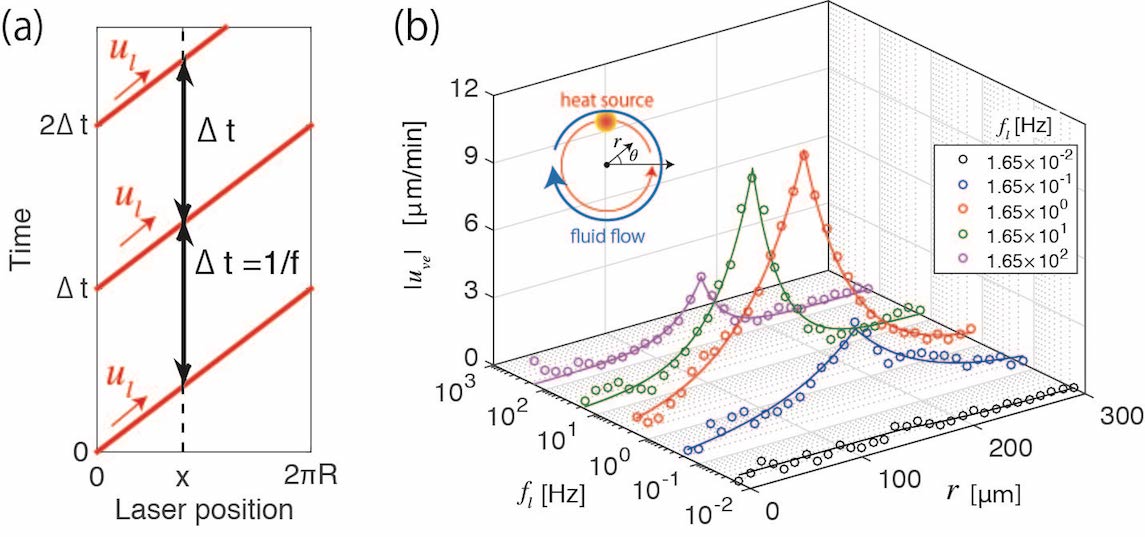}
 \end{center}
 \caption{Frequency-dependent microflow in a moving laser spot along circular path. (a) A moving laser spot of the heat source moves in one direction. The radius of the circular laser path is $R$. (b) Azimuthal component of fluid flow, $\bar{u_{ve}}$, as a function of radial distance $r$ from the center of the circular path.}\label{figS3}
\end{figure}

\section{Theoretical details}
\begin{screen}
\subsection{Thermal expansion of boundary walls by heat spot propagation}
\end{screen}

\begin{figure*}
 \begin{center}
  \includegraphics[width=100mm]{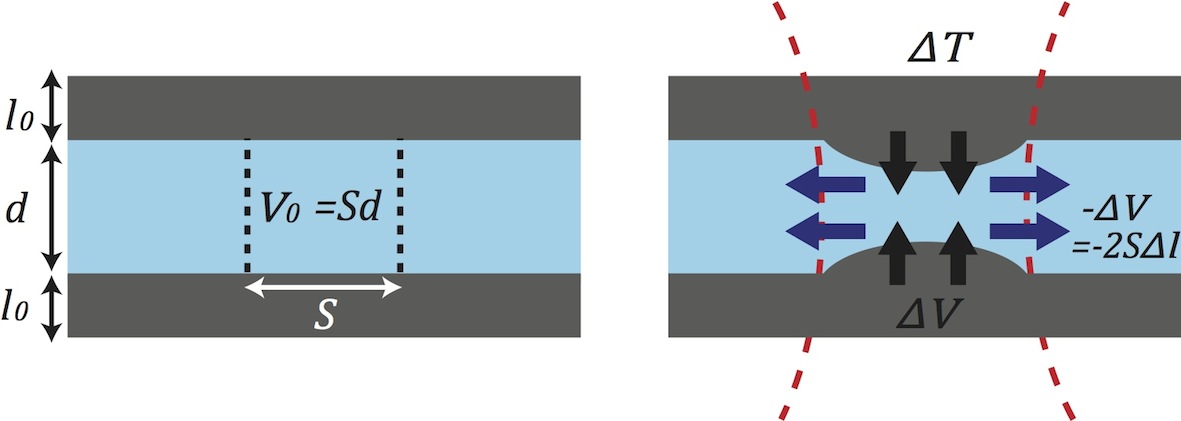}
 \end{center}
 \caption{Schematic illustration of thermal expansion of boundary wall and the compression of enclosed solution.}\label{figS3}
\end{figure*}

The aqueous polymer solution was enclosed with the flexible boundary walls whose substrate is polydimethyl siloxane (PDMS) being deformed upon the exposure of a thermal stimuli. The moving laser spot, that moves at a constant speed $\bm{u_l}=u_l \bm{e_x}=(u_l, 0)$ in $x$ axis, heats water solvent and the temperature of polymer solution $\Delta T(x,y)$ is increased locally. In this occasion, thermal expansion of the solution of 5.0\textit{wt}\% PEG is negligible because of the incompressibility of fluids. The conducted heat thereafter reaches the boundary walls at bottom and ceil and the PDMS substrates thermally expands. increase the temperature of PDMS walls. Thus, the expansion of the flexible boundaries subject to laser heating generates the mechanical force to drive the fluid flow (FIG.S4). The continuum equation for the volume of aqueous solution $V(x,t)$ is
\begin{equation}
\frac{\partial V}{\partial t} + \bm{\nabla} \cdot (V \bm{u}) = 0,
\end{equation}
where $\bm{\nabla} = \frac{\partial}{\partial x}\bm{e_x} + \frac{\partial}{\partial y}\bm{e_y}$ and $\bm{u}=(u,v)$ is the flow induced by thermal expansion of the wall. At the reference frame of the heat spot, we can rewrite the coordinate as $(x', y)=(x-u_{l}t, y)$ and the time-derivative can be rewritten as $\frac{\partial}{\partial t}= \frac{\partial x'}{\partial t} \frac{\partial}{\partial x'}= - u_l \frac{\partial}{\partial x'}= - u_l \frac{\partial}{\partial T} \frac{\partial T}{\partial x'}$. By using this derivative of $T$, the continuum equation of the local volume $V$ in $x$ axis is given by 
\begin{equation}
- u_{l} \frac{\partial V}{\partial T}\frac{\partial T}{\partial x'} + \nabla' \cdot (V (u+u_l)) = 0,
\end{equation}
where $\nabla'$ is $\frac{\partial}{\partial x'}$. Because the motion of a hot spot is symmetric across $x=0$, one can suppose that the change of local volume in $y$ axis does not depend on time and we thus get $V\partial v/\partial y =0$ in order to satisfy the conservation of mass. Accordingly, 
\begin{equation}
- u_{l} \frac{\partial V}{\partial T}\frac{\partial T}{\partial x'} + V \frac{\partial u}{\partial x'} = 0.
\end{equation}
After the exchange of $x$ variable from $x'$ ($\because \partial/\partial x' = \partial x/\partial x' \cdot \partial/\partial x=\partial/\partial x$), we obtain 
\begin{equation}
\frac{\partial u}{\partial x} = u_{l}\frac{\partial (\ln V)}{\partial T}\frac{\partial T}{\partial x}.
\end{equation}

$ $

\begin{screen}
\subsection{Thermo-viscous hydrodynamics}
\end{screen}
In the thin chamber, the momentum balance equation (Stokes equation) is
\begin{equation}
\rho \frac{\partial \bm{u}}{\partial t} = - \nabla p + \nabla \cdot (\eta \nabla \bm{u}) +\bm{K}, \label{Stokes1}
\end{equation}
where $\rho$ is the density of fluid, $\bm{u}=(u,v,w)$ the velocity of fluid flow, $p$ is the pressure, $\eta=\eta(x,y,z)$ is the viscosity of fluid, and $\bm{K}=(K_x, K_y, 0)$ the external force acting on fluid (the friction at the boundary wall)\cite{weinert}. We consider the pressure gradient is zero and viscosity is uniform along $z$ axis, so that the velocity and the viscosity are reduced to $\bm{u}=(u,v)$ and $\eta=\eta(x,y)$, respectively. 

Herein, we first describe the dynamics of fluid flow in a moving thermal field $\Delta T(x',y) = \Delta T(x-u_lt,y)$. As we considered in previous sections, the spot of heating moves at the speed of $\bm{u_l} =(u_l, 0)$. The Stokes equation in $(x',y)$ coordinates yields
\begin{eqnarray}
-\rho u_l \frac{\partial u}{\partial x'}&=& - \frac{\partial p}{\partial x'} + \frac{\partial}{\partial x'} \left(2 \eta \frac{\partial u}{\partial x'} \right) \nonumber \\ 
&+& \frac{\partial }{\partial y} \left[ \eta \left( \frac{\partial u}{\partial y} + \frac{\partial v}{\partial x'} \right) \right] + \eta \frac{\partial^2 u} {\partial z^2} + K_x, \label{stokesX}\\ 
-\rho u_l \frac{\partial v}{\partial x'}&=& - \frac{\partial p}{\partial y} + \frac{\partial}{\partial x'} \left[\eta \left( \frac{\partial u}{\partial y} + \frac{\partial v}{\partial x'} \right)  \right] \nonumber \\ 
 &+& \frac{\partial }{\partial y} \left(2 \eta \frac{\partial v}{\partial y} \right) + \eta \frac{\partial^2 v}{\partial z^2} +K_y.  \label{stokesY}
\end{eqnarray} 
The solution is enclosed in thin chamber with the thickness $d$ so that the friction acts on fluid at the ceiling and bottom walls of $z= \pm d/2$.  For the no-slip boundary condition at the walls, 
\begin{eqnarray}
u(x',y,z)&=&u(x',y,0) \Biggl[1 - \biggl(\frac{2z}{d} \biggr)^2 \Biggr], \\ v(x',y,z)&=&v(x',y,0) \Biggl[1 - \biggl(\frac{2z}{d} \biggr)^2 \Biggr].
\end{eqnarray}
After averaging $u$ and $v$ over $z$ direction ($-d/2 \leq z \leq d/2$), the frictional force in $x'$ and $y$ can be obtained as
\begin{eqnarray}
K_x &=& \frac{12}{d^2} \eta u, \label{fric1} \\
K_y &=& \frac{12}{d^2} \eta v. \label{fric2}
\end{eqnarray}
We then return the coordinates from moving frame $(x',y)$ to laboratory frame $(x,y)$. Substituting $K_x$ and $K_y$ with Eqs.(\ref{fric1}) and (\ref{fric2}), the Stokes equation of Eqs. (\ref{stokesX}) and (\ref{stokesY}) can be rewritten as  
\begin{eqnarray}
0&=& - \frac{\partial p}{\partial x} + \frac{\partial}{\partial x} \left(2 \eta \frac{\partial u}{\partial x} \right) + \frac{\partial }{\partial y} \left[ \eta \left( \frac{\partial u}{\partial y} + \frac{\partial v}{\partial x} \right) \right] \nonumber \\ 
 &+& \eta \frac{\partial^2 u} {\partial z^2} + \frac{12}{d^2} \eta u,\\
0&=& - \frac{\partial p}{\partial y} +  \nonumber + \frac{\partial }{\partial x} \left[ \eta \left( \frac{\partial u}{\partial y} + \frac{\partial v}{\partial x} \right) \right] \frac{\partial}{\partial x} \left(2 \eta \frac{\partial u}{\partial x} \right) \nonumber \\
&+&   \eta \frac{\partial^2 v} {\partial z^2} +\frac{12}{d^2} \eta v.
\end{eqnarray}

However, for the polymer solution enclosed in a micron-scaled channel, the frictional force is dominant rather than viscous drug, such that $\nabla \cdot (\eta \nabla \bm{u}) \ll 12 \eta \bm{u} /d^2$. This yields
\begin{eqnarray}
\frac{\partial p}{\partial x} &=& \frac{12}{d^2} \eta u, \\ \frac{\partial p}{\partial y} &=& \frac{12}{d^2} \eta v.
\end{eqnarray}
The equation sets govern the mechanical balance in $x$ and $y$ directions and then leads to more simple relation as $\nabla \times (\eta \bm{u}) = 0$. Because the viscosity of solution is no longer uniform in the moving thermal gradient, we need to take into account the spatial derivative of $\eta$. However, in $x$ axis for which the horizontal to the heat traveling wave, the gradient of viscosity $\partial \eta/\partial x$ and the gradient of velocity $\partial v/\partial x$ assumed to be small compared to $(\partial \eta/\partial T)(\partial T/\partial y)$ and $\partial u/\partial y$. According to this approximation, the linearization of Stokes equation gives
\begin{equation}
\eta \frac{\partial u}{\partial y} + u \frac{\partial \eta}{\partial T}\frac{\partial T}{\partial y} =0.
\end{equation}
This expression finally allows us to find 
\begin{equation}
\frac{\partial u}{\partial y} =-  u \frac{\partial (\ln \eta)}{\partial T}\frac{\partial T}{\partial y}.
\end{equation}

\begin{screen}
\subsection{Heat conduction in polymer solution}
\end{screen}

\begin{figure*}
 \begin{center}
 \includegraphics[width=100mm]{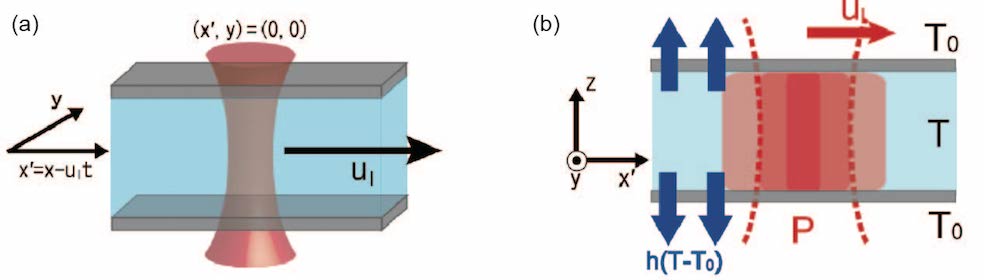}
 \end{center}
 \caption{Schematic illustrations of (a) the reference frame in a moving thermal gradient and (b) the thermal diffusion and sink in the reference frame.}\label{figS4}
\end{figure*}

One of fundamental novelty of this study, compared to previous studies, arises from high thermal insulation of PDMS boundary wall. Its thermal conductivity is 0.15 W/(mK) much smaller than water with 0.59 W/(mK). In this combination of materials, the heat by infrared laser focusing diffuse in an aqueous phase although PDMS wall plays a role of heat sink but it's radiation ability is low. On the other hand, when the aqueous phase is enclosed in a glass with higher thermal conductivity of 1.00 W/(mK) larger than both water and PDMS, the heat transfer is dominantly occurred from water to glass so that the heat dominantly diffuse to $z$ axis, which means the shape of temperature distribution is insensitive to the speed of a moving laser spot. However, heat conduction in lateral direction along the traveling laser spot has to be considered for our experimental system (FIG. S5).

Suppose that the spot of heat source is built in two-dimensional space $\bm{X}=(x,y)$ and moves at constant speed ${u_{l}}$ in $x$ axis. The thermal field due to the moving heat spot is described by the equation of thermal conductivity
\begin{equation}\label{eq:Seq1}
C_{v} \frac{\partial T}{\partial t} - \lambda_h \nabla^2 T = P - h \Delta T,
\end{equation}
where $C_{v}$ is heat capacity and $\lambda_h$ heat conductivity, $P$ is the source of heating spot, $h$ the coefficient of heat sink. The equation is invariant with $\Delta T$ instead of $T$, we can rewrite
\begin{equation}\label{eq:Seq2}
C_{v} \frac{\partial \Delta T}{\partial t} - \lambda_h \nabla^2(\Delta T) = P - h \Delta T.
\end{equation}

In the moving frame of the heat spot, the coordinate and the derivative functions are converted as 
\begin{eqnarray}\label{eq:Seq3}
x' &=& x-u_{l}t \\ 
\frac{\partial}{\partial t} &=& -u_{l} \frac{\partial}{\partial x'} \\
\nabla^2 &=& \frac{\partial^2}{\partial x^2} +\frac{\partial^2}{\partial y^2} =\frac{\partial^2}{\partial x'^2} +\frac{\partial^2}{\partial y^2} = \nabla'^2
\end{eqnarray} 
The shape of laser spot could be given by $P=P_0 \exp[-(x'^2+y^2)/(2b^2)]$ with the spot radius $b$. 

To solve thermal diffusion equation, we perform spatial Fourier transform of temperature difference $\Delta T$ and $P$ by
\begin{eqnarray}\label{eq:Seq4}
\Delta T(x',y) &=& \frac{1}{2\pi} \int^{\infty}_{-\infty} d\bm{k} \hat{T}(\bm{k}) e^{i\bm{k}\cdot \bm{X'}},\\
P(x',y) &=& \frac{1}{2\pi} \int^{\infty}_{-\infty} d\bm{k} P_0 \exp \Biggl(- \frac{b^2 k^2}{2} \Biggr)e^{i\bm{k}\cdot \bm{X'}}.
\end{eqnarray}
By solving thermal diffusion equation after substitution with transformed variables, $\hat{T}(k)$ is given by
\begin{equation}\label{eq:Seq5}
\hat{T}(\bm{k}) = \frac{P_0}{\lambda_h k^2 - i C_{v} \bm{u_{l}} \cdot \bm{k}+ h} \exp \Biggl(- \frac{b^2 k^2}{2} \Biggr).
\end{equation}

We then carried out inverse Fourier transform of $\hat{T}(\bm{k})$ numerically and obtained thermal field at various $u_{l}$ at the moving frame. The unsteady thermal fields are characterized by one physical parameter, the frequency of repetitive thermal stimuli $f_l=u_l/L$. FIG.S5 shows that thermal fields build by the heat spot that moves at $f_l=$0 Hz, 10 Hz, $10^2$ Hz, and $10^3$ Hz. The maximal temperature difference is $\Delta T_{max}=9.6$ K for a steady thermal field. As the speed of moving heat spot increases, the magnitude of $\Delta T$ decreases because the duration time for heating aqueous solution, which is typically $2b/u_{l}$, becomes shorter for larger $u_{l}$. The reduced duration time leads the reduction of heat at arbitrary position.

\begin{figure*}
 \begin{center}
  \includegraphics[width=140mm]{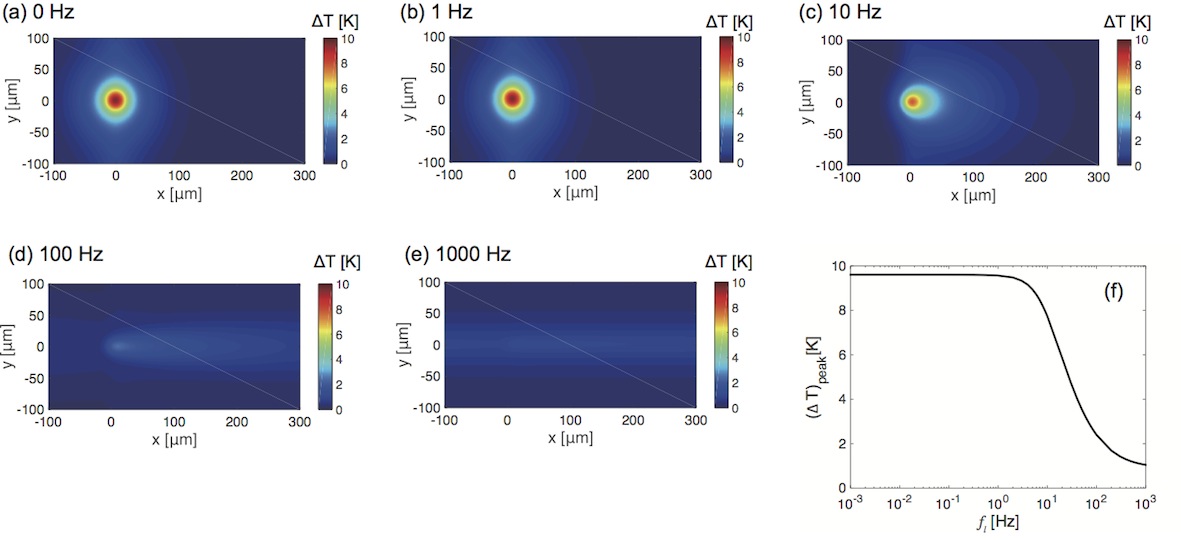}
 \end{center}
 \caption{Numerical simulation of temperature distribution at various speeds of heat wave propagation. Frequency of repetitive thermal stimuli is defined as $f_l=u_l/L$ with the propagation speed of heat wave $u_l$ and the length of scanning path $L$. (a to e) Temperature distribution in the moving frame at $f_l$=(a) 0 Hz, (b) 1 Hz, (c) 10 Hz, (d) 100 Hz, (e) 1000Hz. (f) Maximal temperature difference $\Delta T_{max}$ is plotted as frequency of heat wave.}\label{figS5}
\end{figure*}

$ $

\begin{screen}
\subsection{Heat penetration into PDMS walls}
\end{screen}

A heated polymer solution has the finite temperature difference of $\Delta T=T-T_0$ to the PDMS wall since the temperature of a PDMS wall at the infinity is constant at $T_0$. The diffusion of heat occurs from the polymer solution to the PDMS wall. This heat transfer is important to estimate thermal compression of bulk polymer solution because the depth of this heat conduction decides the strain in the PDMS boundary, as shown later. For this aim, we analyze the penetration depth of transferred heat, which is described by $l_0$, according to thermal diffusion in a PDMS substrate. Thermal field within PDMS $\Delta T'$ is described as below  
\begin{equation}
C_{v}^w \frac{\partial \Delta T'}{\partial t} - \lambda_h^w \nabla_z^2(\Delta T') =  h \Delta T', \label{heatPDMS1}
\end{equation}
where $C_v^w$ is the heat capacity of PDMS, $\lambda_h^w$ the heat conductivity of PDMS, and $h$ is heat transfer coefficient across the surface of PDMS\cite{erickson}\cite{wedershoven}. The term of $h \Delta T'$ represents the transferred heat from the polymer solution, which plays a role of the heat source for PDMS. At the steady state ($\frac{\partial \Delta T'}{\partial t}=0$), since the acquired heat from the polymer solution diffuse through the PDMS substrate, the heat diffusion of $\lambda_h^w \nabla_z^2(\Delta T')$ is comparable to $h \Delta T'$. Approximating $\nabla_z \sim 1/l_0$, we can estimate  the penetration depth of heat into the PDMS wall as
\begin{equation}
l_0 \sim \sqrt{\frac{\lambda_h^w}{h}}. \label{heatPDMS2}
\end{equation}
For PDMS, typical parameters have been known as $\lambda_h^w$=1.5$\times$$10^{-7}$ W/(\SI{}{\micro\meter}$\cdot$K)\cite{erickson} and $h$=5.0$\times$$10^{-10}$ W/(\SI{}{\micro\meter}$^3$$\cdot$K)\cite{wedershoven}, and we can estimate $l_0 \simeq 17 {\rm \mu m}$. This penetration depth of heat is comparable to the thickness of bulk fluids (thickness of the chamber), $d$=\SI{25}{\micro\meter}. 

$ $

\begin{screen}
\subsection{Derivation of Eq. (1) in the main text}
\end{screen}
Thus the associated relations of fluid velocity $u$ are given by
\begin{eqnarray}
\frac{\partial u}{\partial x} &=& u_{l}\frac{1}{V}\frac{\partial V}{\partial T}\frac{\partial T}{\partial x},\\
\frac{\partial u}{\partial y} &=& - u \frac{1}{\eta}\frac{\partial \eta}{\partial T}\frac{\partial T}{\partial y}.
\end{eqnarray}
We define the rate of volume change for temperature $\gamma=(1/V)(\partial V/\partial T)=\partial(\ln V)/\partial T$ and consider that $\gamma$ is constant. In addition, the rate of viscosity change for temperature is defined as $\beta =(1/\eta)(\partial \eta/\partial T)=\partial(\ln \eta)/\partial T$  and assumed as constant as well. 
By substituting $\partial(\ln V)/\partial T$ and $\partial(\ln \eta)/\partial T$ for $\gamma$ and $\beta$, 
\begin{eqnarray}
\frac{\partial u}{\partial x} &=& u_{l}\gamma \frac{\partial T}{\partial x},\\
\frac{\partial u}{\partial y} &=& - u \beta \frac{\partial T}{\partial y}.
\end{eqnarray}

Next, to find the explicit form of thermal expansion coefficient of $\gamma$ for polymer solution, we consider thermal expansion of the PDMS substrate, through the effect of viscoelastic deformation. Suppose that the polymer solution is locally heated by focused infrared laser, the transferred heat from bulk solution induces thermal expansion of soft deformable wall of PDMS. The length of thermal expansion in $z$ axis is assumed to 
\begin{equation}
\Delta l = l_0 \gamma^w \Delta T,
\end{equation} 
where $l_0$ is the penetration depth of transferred heat (Eq.(\ref{heatPDMS2})), and $\gamma^w$ is thermal expansion coefficient of PDMS defined as $\gamma^w=(1/V^w)(\partial V^w/\partial T)$. Because of this thermal expansion, the change of volume in a PDMS channel is $\Delta V^w=2S \Delta l =2S l_0 \gamma^w \Delta T$ with the area of $S$ (FIG. S4).

We describe the local volume of bulk fluid at rest as $V_0=S d$ with the area of $S$ and the thickness of $d$. This bulk fluid is subject to the compression owing to the expanded PDMS wall. Then volume pushed out by PDMS expansion is assumed as $\Delta V =2S \Delta l$. The strain of $\epsilon$ is defined as $\epsilon = - (V-V_0)/V_0 = - \Delta V/V_0$. This expression leads $\gamma=(1/V)(\partial V/\partial T) =(2l_0/d)(1/\epsilon_{\tau})(\partial \epsilon_{\tau}/\partial T) \equiv (2l_0/d) \gamma^w$

To describe the mechanics of local thermal expansion, we employed Voigt model consisted of one spring and one dash-pot. The dynamics of strain relaxation is given by
\begin{equation}
\eta^w \frac{d \epsilon_{\tau}}{dt} + E \epsilon_{\tau} = \sigma, \label{Voigt1}
\end{equation} 
$\epsilon_{\tau}$ is strain, $E$ is the elastic constant, $\eta^w$ is the viscosity of the dash-pot (PDMS substrate), $\sigma$ the stress due to thermal expansion. Because the mechanical stress arises from thermal expansion in the exposure of $\Delta T$, one can yield $\sigma = E \gamma^w \Delta T$ with thermal expansion coefficient of PDMS $\gamma^w$ and Eq. (\ref{Voigt1}) is rewritten as
\begin{equation}
\frac{d \epsilon_{\tau}}{dt} + \frac{1}{\tau} \epsilon_{\tau} = \frac{1}{\tau} \gamma^w \Delta T. \label{Voigt2}
\end{equation}
$\tau=\eta^w/E$ is the characteristic times for viscoelastic relaxation. By solving this differential equation within the frame of one cycle, $t=[n/f_l, (n+1)/f_l]$ ($n=0, 1, 2, \cdots$), the ratio of volume change of solution under compression by expanded walls is
\begin{equation}
\frac{\Delta V}{V_0} = - \gamma \Delta T (1- e^{-\frac{1}{f_l \tau}}),
\end{equation}
where $\Delta V=V-V_0$ and $\gamma =(2l_0/d) \gamma^w$. It then reads the sheathed velocity,
\begin{equation}
\frac{\partial u}{\partial x} = - u_l \gamma (1- e^{-\frac{1}{f_l \tau}}) \frac{\partial T}{\partial x}.
\end{equation}

By taking integral over $x$ and $y$, the fluid velocity eventually yields
\begin{eqnarray}
u &=& - \frac{u_l}{2} \beta \gamma (1- e^{-\frac{1}{f_l \tau}})  (\Delta T)^2.
\end{eqnarray}
The typical penetration depth $l_0$ of PDMS is \SI{17}{\micro\meter} while the depth of microchannel of $d$ is \SI{25}{\micro\meter} in this experiment. We assume that the ratio $2l_0/d$ is close to 1 ($\gamma \approx \gamma^w$), and obtain the Eq. (1) in main text as
\begin{equation}
u_{ve} = - \frac{u_l}{2} \beta \Gamma_{\tau}(\Delta T)^2
\end{equation}
with $\Gamma_{\tau} = \gamma (1- e^{-\frac{1}{f_l \tau}})$.  

$ $

\begin{screen}
\subsection{Reduced viscosity in temperature and solute gradients}
\end{screen}
In a solution of polymer such as polyethylene glycol (PEG), the viscosity of solution $\eta$ depends on both temperature $T$ and the concentration of PEG $c^p$. According to the previous studies, the empirical relation of the viscosity is given by
\begin{equation}
\eta = C \exp\biggl[\frac{B(c^p)}{T-T_c} \biggr],
\end{equation}
where $C$ and $T_c$ are the constants independent of thermodynamic variables $T$ and $c^p$ but $B(c^p)$ is the function showing monotonic increase for $c^p$. This empirical relation has been independently obtained from the study using molecular dynamics simulation\cite{holyst}, such that $\eta =C' \exp[E/(RT)]$ where $C'$ is the constant while $E$ is the variable depending on $c^p$. Given that the small deviations of temperature $\Delta T$ and polymer concentration $\Delta c^p$, the change of viscosity $\Delta \eta$ is described as
\begin{eqnarray}
\Delta \eta &=& \biggl(\frac{\partial \eta}{\partial T} \biggr)_{c^p}\Delta T + \biggl(\frac{\partial \eta}{\partial c^p} \biggr)_{T}\Delta c^p\\
&=& \biggl[- \frac{B}{(T-T_c)^2} \Delta T + \frac{1}{T-T_c} \frac{dB}{dc^p} \Delta c^p \biggr] \eta.
\end{eqnarray}
Then we need to consider how large $1/(T-T_c)$ changes around at room temperature. The previous studies has reported $T_c \approx$ 180 K, suggesting that $1/(T-T_c)$ assumed to be constant at room temperature $T$=\SI{300}{K}. We therefore replace the coefficients of $\Delta T$ and $\Delta c^p$ as constants $\beta_0$ and $\beta_1$ respectively. Eventually, the general form of viscosity change yields
\begin{equation}
\frac{\Delta \eta}{\eta} = - \beta_0 \Delta T + \beta_1 \Delta c^p. \label{viscosity}
\end{equation}

This expression indicates that one needs to consider both temperature and the concentration of polymer in order to capture the change of viscosity in a temperature gradient. To find the explicit form of concentration gradient of polymer, the balance of density flux is considered. Local thermal gradient $\nabla T$ induces the transport of molecules, which named as thermophroesis or the Soret effect. For typical polymer or colloidal particles whose density is heavier than water, thermophoresis depletes these solutes from hot region. The flux of solute due to thermophoresis is described by $J_{Tp}^p=- c^p D_T^p \nabla T$. The spatial distribution of solute concentration is determined by the balance of solute fluxes between thermal diffusion of $J_{Tp}^p$ and normal diffusion $J_{diff}^p=-D^p\nabla c^p$. The phenomenological equation for the net flux $J^p=J_{diff}^p+J_{Tp}^p$ is 
\begin{equation}
J^p=-D^p (\nabla c^p + c^p S_T^p \nabla T),
\end{equation}
where $D^p$ is the diffusion coefficient, $c^p$ the local concentration of the solute and $S_{T}^p$ the Soret coefficient. One can solve this equation in steady state ($J^p=0$), the concentration of the solute yields
\begin{equation}
c^p(r)=c_0^p \exp[-S_T^p \Delta T]
\end{equation}
with the Soret coefficient of $S_T^p=D_T^p/D^p$ and $c_0^p$ is the solute concentration at infinity. 
For small $\Delta T$, we can assume $\Delta c^p \approx -c_0^p S_T^p \Delta T$. By substituting this form in Eq.(\ref{viscosity}), the change of viscosity reads
\begin{equation}
\frac{\Delta \eta}{\eta} = - (\beta_0 + \beta_1 c_0^p S_T^p )\Delta T.
\end{equation}

\begin{figure*}
 \begin{center}
 \includegraphics[width=140mm]{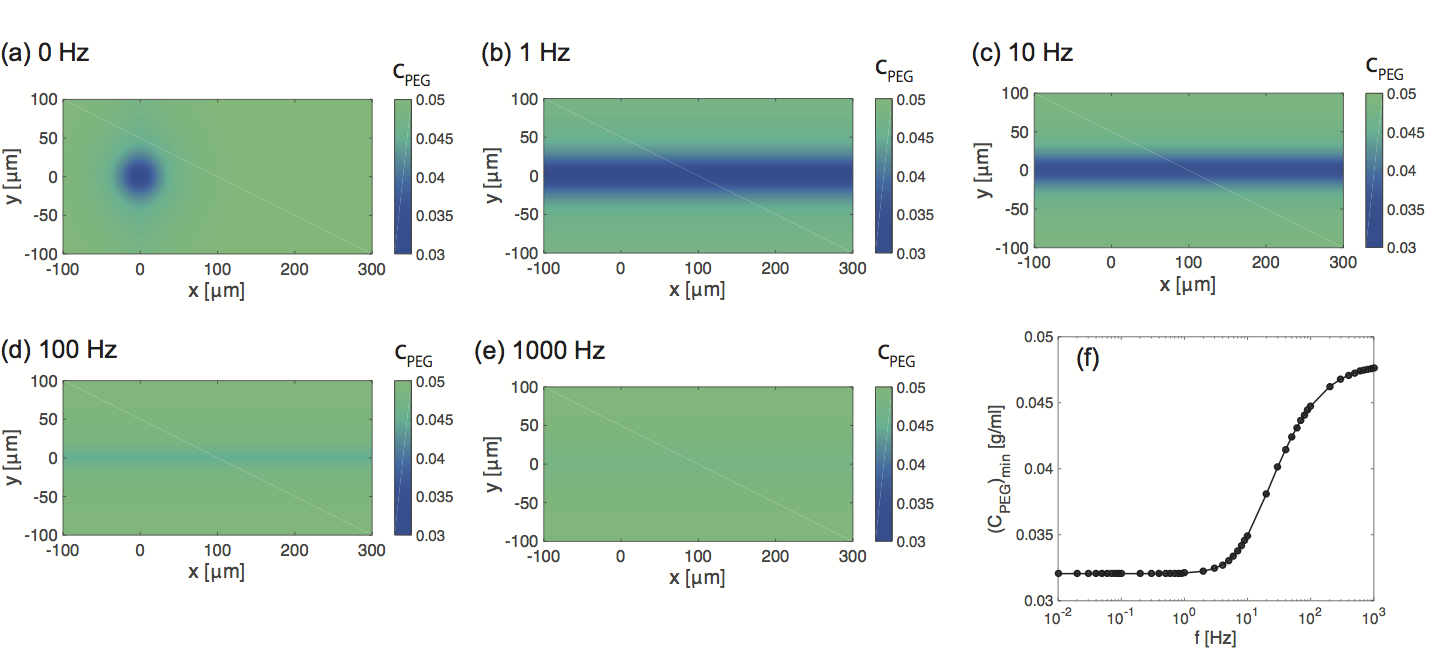}
 \end{center}
 \caption{Numerical simulation of the concentration of PEG, $c^p$, under a moving temperature gradient. The color map of depleted solute from hot region is shown in moving frame of heat wave at various propagation speeds. (a to e) $c^p(x)$ in the moving frame. $f_l$=(a) 0 Hz, (b) 1 Hz, (c) 10 Hz, (d) 100 Hz, (e) 1000 Hz. (f) the minimal concentration of solute versus frequency $f_l$ is plotted.}\label{figS6}
\end{figure*}

$ $

\begin{screen}
\subsection{Thermal molecular focusing}
\end{screen}
Next, we show theoretical details for the derivation of Eq. (8) in the main text by considering the transport equation of DNA molecules that are subject to thermophoresis, diffusiophoresis, and hydrodynamic focusing. The equation of density flux of DNA reads
\begin{equation}
\bm{J}= -D(\bm{\nabla} c + c S_T \bm{\nabla} T) + c \bm{U}_{Dp} + c \bm{U}_{ve}
\end{equation}
where the diffusiophoretic velocity is $\bm{U}_{Dp}=(U_{Dp}, V_{Dp})$ and $V_{Dp}=\frac{k_BT}{3\eta}\lambda^2 c^p (S_T^{p} - 1/T) \nabla_y T$ with the thickness of depletion layer $\lambda$\cite{sano}\cite{mae1}, and the microflow of local hydrodynamic focusing $\bm{U}_{ve} = (U_{ve}, V_{ve})$.  Because $V_{ve}$ is 10 times smaller than $U_{ve}$ in our experiment, we assume that the last term in the equation above is negligible. Hence, as shown in Eq. (3) in the main text, local hydrodynamic focusing (LHF) of $\bm{U}_{ve}=U_{ve}\bm{e}_{x}$ along the path of laser scanning ($x$ axis) is given by
\begin{equation} 
U_{ve}(x) = - U \sinh \Bigl[\frac{x}{f_{l} \tau L} \Bigr] \beta \gamma (\Delta T)^2,
\end{equation}
where $U$ represents frequency-dependence as $U=2L f_{l} \exp[-1/(f_{l}\tau)]$, $\sinh(x/(f_{l} \tau L))$, and $\beta \gamma (\Delta T)^2$ with $\beta=(\beta_0 + \beta_1 c_0^p S_T^p)$ means the multiplicative interplay of thermal viscosity and thermal compression. Since the magnitude of $v_{ve}$ is ~\SI{0.1}{\micro\meter}/sec while the typical length of normal diffusion is $\sqrt{D/f_l} \approx$ \SI{1}{\micro\meter}. Thus the diffusion is relatively faster than the induced flow, meaning that the steady state can be present. By solving the equation in the steady state ($J=0$), the spatial distribution of DNA in $x$ axis yields
\begin{equation}
c(x,y) = c_0 \exp \Biggl[ -S_T \Delta T + V' \Delta c^p + \frac{1}{D} \biggl( \int_{-\infty}^{x} dx^{\prime} U_{ve}(x^{\prime}) \biggr) \Biggr],
\end{equation}
where $V' = 2 \pi a \lambda^2$ with the radius of gyration of DNA of $a$. We can renormalize the DNA concentration relative to the accumulated amount in the absence of the microflow as
\begin{equation}
\frac{c(x)}{c_0 \exp[ -S_T \Delta T + V' \Delta c^p]} =\exp \Biggl[\frac{1}{D} \biggl( \int_{-\infty}^{x} dx^{\prime} U_{ve}(x^{\prime})\biggr)\Biggr].
\end{equation}
We define this expression as $A(x)$ in order to examine the index of molecular focusing. 

To analytically solve $A(x)$, we need to derive the explicit form of $U_{ve}(x)$. For a heat spot that moves in the path $-L \leq x \leq L$ at $y=0$, the spatial distribution of temperature increase, $\Delta T(x,y)$, follow $\Delta T(x,0) \exp \bigl(- \frac{y^2}{2w^2} \bigr)$, with the width of temperature distribution $w$. We then consider the microflow on the path of the moving heat spot ($y=0$). On the one hand, the temperature gradient that moves in forward direction drives the microflow $u_{ve}^f$ at the position $x$ ($-L \leq x \leq L$) as
\begin{equation}
u_{ve}^f(x)=  - \frac{u_l}{2} (\beta_0 + \beta_1 c_0^p S_T^p) \gamma (1- e^{-\frac{L+x}{f_l\tau L}}) (\Delta T(x,0))^2. \label{forwardflow}
\end{equation}
On the other hand, when the heat spot moves in opposite direction (backward), the microflow $u_{ve}^b$ at the position $x$ reads
\begin{equation}
u_{ve}^b(x)=  + \frac{u_l}{2} (\beta_0 + \beta_1 c_0^p S_T^p) \gamma (1- e^{-\frac{L-x}{f_l\tau L}}) (\Delta T(x,0))^2. \label{backflow}
\end{equation}

By summing reciprocal microflows of Eqs. (\ref{forwardflow}) and (\ref{backflow}), we obtain the expression of LHF as
\begin{equation}
U_{ve}(x)=  - u_l e^{-\frac{1}{f_l \tau}} \sinh \Bigl[\frac{x}{f_l \tau L} \Bigr] (\beta_0 + \beta_1 c_0^p S_T^p) \gamma  (\Delta T(x,0))^2.
\end{equation}

 $ $

\begin{table*}[h]
  \begin{tabular}{|l|c|l|} \hline
    Parameter & Symbol & Number \\ \hline \hline
   Half of the length of laser scanning line & $L$ & \SI{80}{\micro\meter} \\
   Relaxation time constant of PDMS & $\tau$ & \SI{1.2}{sec} \\
    Heat capacity of water & $C_v$  & 4.2$\times$$10^{-12}$ J/(\SI{}{\micro\meter}$^3$$\cdot$K)\cite{erickson}\\
    Heat conductivity of water & $\lambda_h$ & 6.1$\times$$10^{-7}$ W/(\SI{}{\micro\meter}$\cdot$K)\cite{erickson} \\ 
Heat conductivity of PDMS & $\lambda_h^w$ & 1.5$\times$$10^{-7}$ W/(\SI{}{\micro\meter}$\cdot$K)\cite{erickson} \\    
Heat transfer coefficient & $h$  & 5.0$\times$$10^{-10}$ W/(\SI{}{\micro\meter}$^3$$\cdot$K) \cite{wedershoven} \\
   Thermo-viscous coefficient of water & $\beta_0$  & 2.2$\times$$10^{-2}$ 1/K  \\
   The rate of viscosity change by PEG conc. & $\beta_1$  & 27.7 ml/g \\
   Thermo-expansion coefficient of PDMS & $\gamma$  & 3.1$\times$$10^{-4}$ 1/K  \\
   The size of a moving heat spot & $b$ & \SI{7.5}{\micro\meter} \\
   Soret coefficient of PEG20000 & $S_T^p$  & 8.89$\times$$10^{-2}$ 1/K \cite{chan}\\
   Soret coefficient of DNA & $S_T$  & 3.80$\times$$10^{-1}$ 1/K \cite{mae1}\\
   The depth of depletion layer & $\lambda$  & 2.5 nm \\
   The gyration radius of DNA & $a$  & \SI{0.1}{\micro\meter}  \\
   Diffusion coefficient of PEG20000 & $D^p$ & \SI{58}{\micro\meter}$^2$/s  \\
   Diffusion coefficient of DNA & $D$  & \SI{2.89}{\micro\meter}$^2$/s \cite{mae1}\\ \hline
  \end{tabular}
   \caption{The list of parameters used in this study for numerical simulation}\label{TableS1}
\end{table*}


\begin{thebibliography}{45}
\bibitem{supporting2} The concentration of PEG yields $c^p=c_0^p \exp[-S_T^p \Delta T]$ where $c_0^p$ is PEG concentration at infinity.
\end{thebibliography}
\end{document}